\documentclass[twocolumn,twocolappendix]{aastex631}
\usepackage{CJK}
\usepackage{graphicx}
\usepackage{amsmath}
\usepackage{physics}
\usepackage{cancel}
\usepackage[version=4]{mhchem}

\graphicspath{{./}{figures/}}

\newcommand{\E}{\mathcal{E}}
\newcommand{\F}{\mathbf{F}_{r}}
\newcommand{\G}{G^{0}}
\newcommand{\vel}{\mathbf{v}}
\newcommand{\kp}{\kappa_{\rm{P}}}
\newcommand{\kr}{\kappa_{\rm{R}}}
\newcommand{\eg}{e_{\text{g}}}
\newcommand{\tg}{T_{\text{g}}}
\newcommand{\arad}{\mathbf{a}_{\text{rad}}}

\newcommand{\rin}{r_{\rm{in}}}
\newcommand{\rout}{r_{\rm{out}}}
\newcommand{\cmsg}{cm$^{2}\cdot$g$^{-1}$}
\newcommand{\gcmc}{g$\cdot$cm$^{-3}$}
\newcommand{\M}{M_{\star}}

\defcitealias{chen2024}{CI24}

\submitjournal{ApJ}
\shorttitle{From CEE to LRNe I}
\shortauthors{Zhuo Chen}

\begin{document}
\begin{CJK*}{UTF8}{gbsn}
\title{From Common Envelope Evolution to Luminous Red Novae I: A One-dimensional Radiation Hydrodynamic Model}

\correspondingauthor{Zhuo Chen (陈卓)}
\email{chenzhuo\_astro@mail.tsinghua.edu.cn}\\

\author[0000-0001-7420-9606]{Zhuo Chen (陈卓)}
\affiliation{Institute for Advanced Study, Tsinghua University \\
Beijing 100084, China}

\begin{abstract}
The acceleration and unbinding of the common envelope during the plunge-in phase are governed by complex physical processes that often manifest observationally as luminous red novae. We investigate the dynamics of this phase using one-dimensional radiation hydrodynamic simulations evolved with the code {\tt Guangqi}. We perform a parameter survey to quantify the impact of key physical conditions on the unbound mass fraction, $\eta$, and the resulting light curves. Our survey spans a range of radiation-to-gas internal energy ratios ($\mathcal{E}/e_{\text{g}}\in[0.2,3.2]$), ratios of total envelope energy to gravitational binding energy ($\zeta\in[0.54,2.87]$), and mass injection rates ($\dot{M}\in[2.5,10]M_{\odot}/\rm{yr}$), while covering both subsonic and supersonic expansion regimes ($v_{\rm ej}/v_{\rm esc}\in[0.3,0.6]$). We demonstrate that: (1) radiation pressure becomes the dominant driver of mass ejection in the high-opacity, high-luminosity region immediately below the recombination front; (2) $\eta$ exhibits a nonlinear dependence on $\zeta$, which is modulated by the mass injection rate and gravitational potential; and (3) the recombination of atomic to molecular hydrogen ($\ce{H}\to\ce{H2}$) releases latent heat that sustains a secondary plateau in the late-time light curve. These findings are substantiated by detailed error analysis and convergence testing presented in the Appendices.
\end{abstract}


\section{Introduction}

Common envelope evolution (CEE) is one of the most poorly understood stages in binary evolution \citep{ivanova2013}, owing to the rarity of these events, the challenges of direct observation \citep{tylenda2011,blagorodnova2017,cai2019,pastorello2019,blagorodnova2021,pastorello2021a,pastorello2021b}, and its inherent multi-physics complexity. During this phase, the binary cores' orbital energy is converted to kinetic, thermal, magnetic, and radiation energy. If the orbital energy is deposited in a highly optically thick environment in a short time---such as during the plunge-in phase---radiation pressure may exceed the gas pressure, promoting envelope ejection. After the CEE, the separation between the stellar cores can decrease by up to two orders of magnitude. Depending on the progenitor system, the resulting binaries may subsequently evolve into gravitational wave sources \citep{dominik2012}, double neutron stars \citep{chattarj2026}, X-ray binaries \citep{kalogera1998,xing2025}, or progenitors of type Ia supernovae \citep{liu2023}. Unfortunately, the gap in our understanding connecting the CEE phase to its final products remains significant; consequently, we cannot yet establish a definitive one-to-one correspondence between binary progenitors and the resulting evolved binaries \citep{han2020,chenxuefei2024}.

Numerous three-dimensional (3D) hydrodynamic \citep{ricker2012,nandez2014,nandez2015,ivanova2016,nandez2016,ohlmann2016,chamandy2018,iaconi2019,prust2019,sand2020,moreno2022,lau2022a,lau2022b,chamandy2024} and magnetohydrodynamic (MHD) \citep{ondratschek2022,vetter2024,vetter2025} simulations have been carried out to study the final separation problem together with the envelope unbinding problem. Typically, these simulations are performed under adiabatic assumption, as fully modeling convective and radiative energy transport in 3D is computationally prohibitive (though see \citet{goldberg2022} and \citet{lau2025}). In such adiabatic models, envelope unbinding is primarily ascribed to the pressure gradients and angular momentum transport via turbulence or magnetic fields. However, neglecting radiation transport is a significant simplification. While the adiabatic assumption may be appropriate for a small, low-luminosity envelope early in the interaction, it inevitably breaks down as the envelope expands and becomes more luminous. The recent progress made by \citet{lau2025} shows that the radiation transport inhibits the ejection of the envelope, compared to the adiabatic simulation, because radiation transport is an effective way to dissipate energy in the optically thin region. On the other hand, radiation pressure is crucial in driving the winds of Wolf-Rayet \citep{grafener2017,sander2020} and AGB stars \citep{hofner2018,decin2021}. Finally, including radiation transport is essential for predicting the observational signatures of these events \citep{hatfull2021,hatfull2025} and bridging the gap between theory and observation \citet{chen2024} (hereafter \citetalias{chen2024}).

Alternatively, one-dimensional (1D) models have been developed to resolve the microphysics of CEE, focusing on recombination energy, convective and radiative energy transport. Studies suggest that recombination energy may only play a limited role in envelope unbinding \citep{grichener2018,soker2018}. The role of convection, however, appears more critical. A convective envelope can efficiently transport energy outward, allowing the secondary star to spiral inward \citep{wilson2020,wilson2022,noughani2026}. This inspiral may be halted if the companion encounters a deep, stable radiative layer, a process that could ultimately determine the final binary separation \citep{hirai2022}. Detailed 1D simulations, such as a study of a neutron star and a supergiant system by \citet{fragos2019}, provide a framework for exploring these intricate processes. A key remaining uncertainty in the overall physical process is the role of radiation transport. When the radiation flux and opacity are high, radiation pressure can become a dominant force capable of driving an outflow.

\begin{figure}
    \centering
    \includegraphics[width=\linewidth]{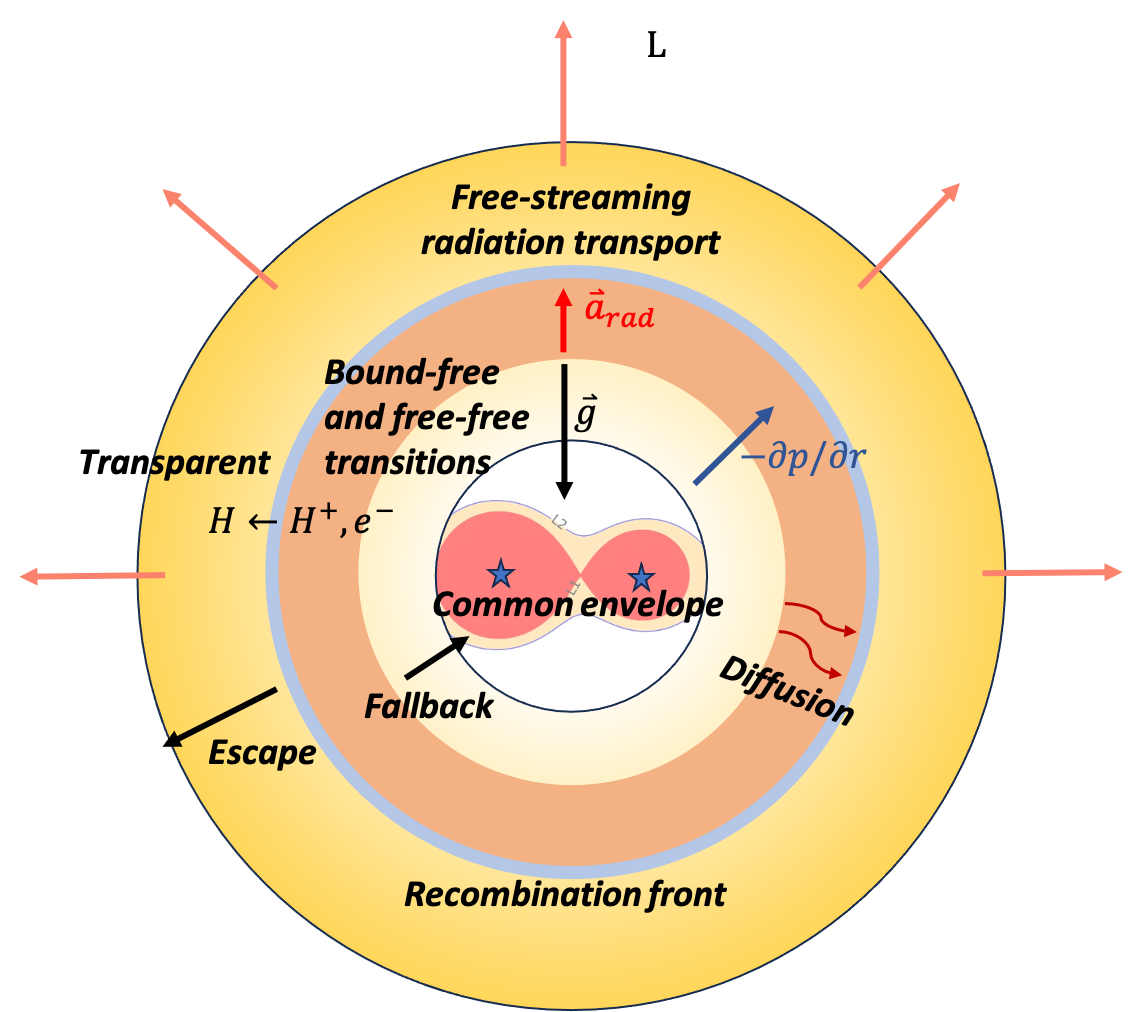}
    \caption{Schematic illustration of relevant physical processes in CEE and Luminous Red Novae (LRNe) implemented in our model. The binary cores drive an expanding envelope, which is decelerated by gravity and accelerated by radiation and gas pressure gradients. A portion of the envelope falls back onto the binary, while the remainder escapes the system. A recombination front divides the expanding envelope into an opaque ionized inner region and a transparent neutral outer region. Radiation diffuses within a shallow layer of the ionized region before transiting to a free-streaming state in the neutral region.}
    \label{fig:schematic}
\end{figure}

Figure \ref{fig:schematic} outlines some relevant physical processes that we have noted in \citepalias{chen2024}. In this work, we improved the 1D radiation hydrodynamic (RHD) model by incorporating radiation pressure self-consistently compared to \citepalias{chen2024}. We survey some key parameters and examine their influence on the unbound mass fraction $\eta$ and observable imprints. The paper is organized as follows. Section \ref{sec:phymodel} elucidates our physical framework, including the governing RHD equations, initial and boundary conditions, and expanding envelope models. Section \ref{sec:results} presents and analyzes our simulation results. Finally, we summarize our conclusions and suggest future research directions in Section \ref{sec:conclusion}.

\section{Physical model}\label{sec:phymodel}

\subsection{The governing equations}

We solve the RHD equations in the laboratory frame with flux-limited diffusion (FLD) approximation and a mixture of hydrogen and helium gas species with {\tt Guangqi} \citep{chen2025}. Due to the presence of a tensor product, for compactness and generality, we write down the governing equations in the vector form, while solving them in spherical coordinates with spherical symmetry:
\begin{eqnarray}
    \pdv{\rho}{t}+\nabla\cdot(\rho\vel)&=&0,   \label{eqn:mass}\\
    \pdv{\rho\vel}{t}+\nabla\cdot(\rho\vel\vel+p\mathbb{I})&=&\rho(\arad-\nabla\phi),   \label{eqn:mom}\\
    \pdv{E}{t}+\nabla\cdot[(E+p)\vel]&=&\rho\vel\cdot(\arad-\nabla\phi)+\G,   \label{eqn:hydroenergy}\\
    \pdv{\E}{t}+\nabla\cdot(\F+\E\vel)&=&-\G-\mathcal{P}_{r}:\nabla\vel,    \label{eqn:radenergy}    \\
    E&=&\rho\frac{|\vel|^2}{2}+\eg(\rho,\tg),    \label{eqn:eos}\\
    \G&=&\kp\rho c(\E-a_{\rm{r}}\tg^{4})
\end{eqnarray}
where $\rho,\vel=[v_{r},0,0],p,\tg,\eg$, and $E$ are the density, velocity, pressure, temperature, internal energy, and total energy of the gas, respectively. The radiation and gravitational force accelerations are denoted by $\arad$ (defined later) and $-\nabla\phi=[-g,0,0]$, where $\phi=-G\M/r$ is the gravitational potential, and $\M$ and $G$ are the mass of the central object and gravitational constant. The gas and radiation exchange heat at the rate of $\G$, in which $\kp$ is the Planck mean opacity in the co-moving frame.

We adopt the FLD radiation transport model \citep{levermore1981,levermore1984}, which provides the closure relations between the radiation energy density $\E$, radiation flux $\F$, and radiation pressure tensor $\mathcal{P}_r$,
\begin{eqnarray}
    \F&=&-\frac{c\lambda(\mathcal{R})}{\kr\rho}\nabla\E, \\
    \mathcal{P}_{r}&=&\frac{\E}{2}[(1-f)\mathbb{I}+(3f-1)\mathbf{n}\mathbf{n}],\label{eqn:Pr}
\end{eqnarray}
where $\kr$, $c$, $\mathbf{n}=\F/|\F|$ are the Rosseland mean opacity in the co-moving frame, speed of light, and the unit vector in the direction of $\F$. The aforementioned radiation acceleration term is calculated by $\arad=\kr\F/c=[a_{\rm{rad}},0,0]$. With dimensional analysis \citep{colombo2019}, it can be shown that,
\begin{equation}\label{eqn:approximation1}
    -\nabla\cdot\mathcal{P}_{r}\approx\frac{\rho\kr\F}{c}=\rho\arad,
\end{equation}
when $v_{\text{gas}}/c\ll1$ and $\F$ is not varying rapidly. Thus, we can equivalently say that gas is accelerated by radiation pressure. This approximation is useful in examining the energy conservation of our simulations (see Appendix \ref{app:energyconservation}). Meanwhile, the flux limiter $\lambda$ and other two dimensionless variables $\mathcal{R}$ and $f$ are calculated by,
\begin{eqnarray}
    \lambda(\mathcal{R})&=&\frac{2+\mathcal{R}}{6+3\mathcal{R}+\mathcal{R}^2},   \\
    \mathcal{R}&=&\frac{|\nabla\E|}{\kr\rho\E}, \\
    f&=&\lambda+\lambda^{2}\mathcal{R}^{2}.
\end{eqnarray}
These variables make $\F$ and $\mathcal{P}_{r}$ have the following desired asymptotic behaviors with smooth transitions in between,
\begin{equation}\label{eqn:asymptotic1}
    \F\rightarrow\begin{cases}
	c\nabla\E/(3\kr\rho)&\quad\quad\rm{optically\ thick,	}\\
	c\E\mathbf{n} &\quad\quad\rm{optically\ thin,}\\
\end{cases}
\end{equation}
and
\begin{equation}\label{eqn:asymptotic2}
    \mathcal{P}_{r}\rightarrow\begin{cases}
	\E\mathbb{I}/3&\quad\quad\rm{optically\ thick,	}\\
	\E\mathbf{n}\mathbf{n} &\quad\quad\rm{optically\ thin.}\\
\end{cases}
\end{equation}

We adopt the same opacity tables from \citetalias{chen2024}, which combine dust \citep{semenov2003} and gas opacities \citep{malygin2014}. Equation \ref{eqn:eos} describes the total energy, which includes the recombination energies by incorporating the internal energy through a general equation of state (EoS) $\eg(\rho,\tg)$ consisting of \ce{H2}, \ce{H}, \ce{H+}, \ce{He}, \ce{He+}, \ce{He^2+}, and \ce{e-} (see Appendix A of \citetalias{chen2024}). The adopted hydrogen and helium mass fraction are $X=0.74$ and $Y=0.26$.

In {\tt Guangqi}, the hydrodynamic equations \ref{eqn:mass}-\ref{eqn:hydroenergy} (without $\G$) are solved by a general EoS HLLC Riemann solver \citep{chen2019} with MUSCL scheme and operator unsplit method \citep{chen2025}. We set the CFL number to be 0.5.

The radiation and gas energy are coupled in thermodynamics, described by the following thermodynamic and radiation transport equations,
\begin{eqnarray}
    \pdv{\eg}{t}&=&\G, \label{eqn:gasthermodynamics}\\
    \pdv{\E}{t}+\nabla\cdot[(\F+\E\mathbf{v})]+\mathcal{P}_{r}:\nabla\mathbf{v}&=&-\G. \label{eqn:radthermodynamics}
\end{eqnarray}
Equation \ref{eqn:gasthermodynamics} and \ref{eqn:radthermodynamics} are integrated implicitly by the GMRES Krylov iterative solver with {\tt Petsc} \citep{petsc-efficient,petsc-user-ref}. We adopt a relative tolerance of error $\epsilon_{r}=10^{-10}$. The thermodynamic system usually has a shorter characteristic timescale than the hydrodynamic system. Thus, to accurately integrate the thermodynamic system after each integration of the hydrodynamic system, we divide the hydrodynamic timestep into $n_{\rm{sub}}=6$ substeps, with each sub-timestep increases by a ratio of $q_{t}=1.4$, i.e., $\delta t_{m+1}=q_{t}\delta t_{m}$, where $m$ is the substep number.

\subsection{Computational domain and numerical setups}

Informed by recent CEE simulations of massive binaries \citep{fragos2019,vetter2024,lau2025}, we define the computational domain as $[\rin,\rout]$, where $\rin=450\,R_{\odot}$ and $\rout=1.5\times10^{4}\,R_{\odot}$. We assume a central gravitational source with a fixed mass $\M=15\,M_{\odot}$ at the origin, representing the enclosed mass of the binary and the inner common envelope. The binary cores are assumed to orbit at radii $r < \rin$, depositing energy into the envelope and imparting it with initial kinetic and thermal energy. Varying $\M$ and $\rin$ reflects the evolutionary stage of the progenitors and can alter the depth of the gravitation potential, a factor that remains independent of the hydrogen/helium recombination energy. Consequently, adopting different values for $\M$ and $\rin$ may, in principle, yield different quantitative results; we therefore anticipate that a more thorough investigation incorporating the specific binary evolution stages will be necessary in future work.

The computational grid has a base resolution of $N_{x}=2048$, with one level of mesh refinement covering the region $[\rin,2\rin]$. We employ a stretched coordinate system where the cell size increases geometrically, such that $\Delta r_{i+1}=1.0021\Delta r_{i}$, where $i$ is the cell index. The initial background conditions are set to a density profile of $\rho(r)=\rho_0(\rin/r)^{1.5}$ (with $\rho_0=10^{-17}$\gcmc), a radial velocity $v_{r}=0$, and a gas temperature $\tg=200$K. The initial mass and energy of this background medium are negligible compared to the expanding envelope, ensuring they impose minimal impact on the resulting dynamics.

\subsection{Boundary conditions}\label{sec:boundcond}

We assume free-streaming conditions at the outer boundary, described by:
\begin{eqnarray}
\pdv{(\E r^2)}{r}=0.
\end{eqnarray}
To numerically enforce the optically thin assumption, we constrain the optical depth in the ghost cells by setting $(\rho\kr)_{N_{x}+\frac{1}{2}}=\min\{(\rho\kr)_{N_{x}},1/(3\rout)\}$. Here, the subscripts $N_{x}$ and $N_{x}+\frac{1}{2}$ denote values at the outermost computational cell and the outer boundary surface, respectively. This constraint guarantees that the photon mean free path at the boundary is at least three times the domain radius. Consequently, the radiation pressure tensor approximates the free-streaming limit, $\mathcal{P}_{r}\approx\E\hat{r}\hat{r}$. For the gas, we apply a standard outflow boundary condition: hydrodynamic variables are copied into the ghost cells, with the radial velocity constrained to $v_{r}=\max\{v_{r},0\}$ to prevent spurious inflow.

The envelope expands into the computational domain through the inner boundary. During the expanding phase, we set the inner boundary conditions for radiation and hydrodynamics to match the envelope's physical properties, including: $\rho,\mathbf{v},\E$, and $\tg$, and $\E=a_{\rm{r}}\tg^{4}$ for the inner boundary. After the expanding phase, we switch the inner boundary to an inflow-only boundary condition. During this time, the ghost cell copies all radiation and hydrodynamic variables from the innermost computational cell, and we impose $v_{r}=\min\{v_{r},0\}$ to accommondate possible fallback gas. While this simple inner boundary allows gas to flow through the inner boundary, it is most suitable to supersonic flows. Applying it to the subsonic regime may introduce more numerical errors. In a post-processing examination, we confirm that the subsonic state in most simulations is short ($\sim10$ days) compare to the envelope injection time (50 days) and the error and convergence analysis (Appendices \ref{app:energyconservation} and \ref{app:convergence}) show that most simulations are accurate and converged. Throughout the simulations, we maintain $\partial\E/\partial r=0$ at the inner boundary, thus, $\F=0$ at the inner boundary. We also find that the inner boundary is always in an optically thick state, and $\mathcal{P}_{r}\approx\E\mathbb{I}/3$ at the inner boundary.

\subsection{Expanding envelope models}

\begin{table*}[]
    \centering
    \begin{tabular}{cccc|cc|cccccc|cc}\hline\hline
    \multicolumn{4}{c|}{common}  &   \multicolumn{2}{c|}{$\dot{M}=2.5\frac{M_{\odot}}{\rm{yr}}$}    & \multicolumn{6}{c|}{$\dot{M}=5\frac{M_{\odot}}{\rm{yr}}$}   &   \multicolumn{2}{c}{$\dot{M}=10\frac{M_{\odot}}{\rm{yr}}$} \\\hline
    model &   $\frac{\E}{\eg}$  &   $\bar{v}_{\rm{ej}}$   &   $\zeta_1$      &   $\zeta$      &   $\mathcal{M}$   &   $\rho_{\rm{ej}}$    &   $T_{\rm{g}}$    &  $\dot{e_{k}}$ & $\dot{\E}+\dot{\eg}$ &  $\zeta$    &    $\mathcal{M}$    &   $\zeta$     &       $\mathcal{M}$    \\
    & &   &  &           &     &   (\gcmc)   &   (K)   &   ($\rm{erg}\cdot\rm{s}^{-1}$) & ($\rm{erg}\cdot\rm{s}^{-1}$) &     &   & &              \\
    (1) &   (2) &   (3) &   (4) &   (5) &   (6) &   (7) &   (8) &   (9) &   (10) &   (11)   & (12) & (13) & (14) \\\hline
    E02v03  &   0.2 & 0.3 & 0.09 & 0.54 & 1.22 & $7.57\times10^{-9}$ & $4.78\times10^{4}$ & $1.81\times10^{39}$ & $9.86\times10^{39}$ & 0.58 & 1.03 & 0.62 & 0.93 \\
    E02v04  &   0.2 & 0.4 & 0.16 & 0.58 & 1.78 & $5.67\times10^{-9}$ & $4.41\times10^{4}$ & $3.21\times10^{39}$ & $9.55\times10^{39}$ & 0.64 & 1.46 & 0.67 & 1.29 \\
    E02v05  &   0.2 & 0.5 & 0.25 & 0.65 & 2.35 & $4.54\times10^{-9}$ & $4.14\times10^{4}$ & $5.02\times10^{39}$ & $9.29\times10^{39}$ & 0.71 & 1.93 & 0.75 & 1.67 \\
    E02v06  &   0.2 & 0.6 & 0.36 & 0.74 & 2.90 & $3.78\times10^{-9}$ & $3.93\times10^{4}$ & $7.22\times10^{39}$ & $9.04\times10^{39}$ & 0.81 & 2.44 & 0.85 & 2.06 \\
    E04v03  &   0.4 & 0.3 & 0.09 & 0.66 & 1.03 & $7.57\times10^{-9}$ & $5.80\times10^{4}$ & $1.81\times10^{39}$ & $1.25\times10^{40}$ & 0.71 & 0.93 & 0.77 & 0.84 \\
    E04v04  &   0.4 & 0.4 & 0.16 & 0.72 & 1.44 & $5.67\times10^{-9}$ & $5.35\times10^{4}$ & $3.21\times10^{39}$ & $1.20\times10^{40}$ & 0.76 & 1.29 & 0.81 & 1.17 \\
    E04v05  &   0.4 & 0.5 & 0.25 & 0.79 & 1.89 & $4.54\times10^{-9}$ & $5.03\times10^{4}$ & $5.02\times10^{39}$ & $1.17\times10^{40}$ & 0.84 & 1.66 & 0.89 & 1.50 \\
    E04v06  &   0.4 & 0.6 & 0.36 & 0.89 & 2.38 & $3.78\times10^{-9}$ & $4.78\times10^{4}$ & $7.22\times10^{39}$ & $1.15\times10^{40}$ & 0.93 & 2.05 & 0.98 & 1.85 \\
    E08v03  &   0.8 & 0.3 & 0.09 & 0.89 & 0.93 & $7.57\times10^{-9}$ & $7.05\times10^{4}$ & $1.81\times10^{39}$ & $1.75\times10^{40}$ & 0.96 & 0.84 & 1.05 & 0.76 \\
    E08v04  &   0.8 & 0.4 & 0.16 & 0.93 & 1.29 & $5.67\times10^{-9}$ & $6.50\times10^{4}$ & $3.21\times10^{39}$ & $1.68\times10^{40}$ & 1.00 & 1.17 & 1.08 & 1.06 \\
    E08v05  &   0.8 & 0.5 & 0.25 & 1.00 & 1.66 & $4.54\times10^{-9}$ & $6.10\times10^{4}$ & $5.02\times10^{39}$ & $1.64\times10^{40}$ & 1.07 & 1.50 & 1.14 & 1.36 \\
    E08v06  &   0.8 & 0.6 & 0.36 & 1.10 & 2.04 & $3.78\times10^{-9}$ & $5.80\times10^{4}$ & $7.22\times10^{39}$ & $1.60\times10^{40}$ & 1.16 & 1.85 & 1.23 & 1.68 \\
    E16v03  &   1.6 & 0.3 & 0.09 & 1.35 & 0.84 & $7.57\times10^{-9}$ & $8.59\times10^{4}$ & $1.81\times10^{39}$ & $2.79\times10^{40}$ & 1.48 & 0.76 & 1.64 & 0.69 \\
    E16v04  &   1.6 & 0.4 & 0.16 & 1.37 & 1.17 & $5.67\times10^{-9}$ & $7.91\times10^{4}$ & $3.21\times10^{39}$ & $2.67\times10^{40}$ & 1.49 & 1.06 & 1.64 & 0.96 \\
    E16v05  &   1.6 & 0.5 & 0.25 & 1.43 & 1.50 & $4.54\times10^{-9}$ & $7.42\times10^{4}$ & $5.02\times10^{39}$ & $2.59\times10^{40}$ & 1.54 & 1.36 & 1.68 & 1.23 \\
    E16v06  &   1.6 & 0.6 & 0.36 & 1.51 & 1.85 & $3.78\times10^{-9}$ & $7.05\times10^{4}$ & $7.22\times10^{39}$ & $2.53\times10^{40}$ & 1.62 & 1.68 & 1.75 & 1.52 \\
    E32v03  &   3.2 & 0.3 & 0.09 & 2.34 & 0.76 & $7.57\times10^{-9}$ & $1.05\times10^{5}$ & $1.81\times10^{39}$ & $5.03\times10^{40}$ & 2.60 & 0.69 & 2.92 & 0.62 \\
    E32v04  &   3.2 & 0.4 & 0.16 & 2.31 & 1.06 & $5.67\times10^{-9}$ & $9.66\times10^{4}$ & $3.21\times10^{39}$ & $4.80\times10^{40}$ & 2.55 & 0.96 & 2.85 & 0.86 \\
    E32v05  &   3.2 & 0.5 & 0.25 & 2.34 & 1.36 & $4.54\times10^{-9}$ & $9.05\times10^{4}$ & $5.02\times10^{39}$ & $4.63\times10^{40}$ & 2.56 & 1.23 & 2.84 & 1.12 \\
    E32v06  &   3.2 & 0.6 & 0.36 & 2.39 & 1.68 & $3.78\times10^{-9}$ & $8.59\times10^{4}$ & $7.22\times10^{39}$ & $4.51\times10^{40}$ & 2.61 & 1.52 & 2.87 & 1.38 \\
    \hline\hline
    \end{tabular}
    \caption{Columns 1 to 4 are the common parameters of the three groups of models with different mass expanding rates $\dot{M}$. They are the model name, radiation to gas internal energy ratio $\E/\eg$, envelope's initial speed to the escape speed ratio $\bar{v}_{\rm{ej}}$, kinetic to gravitational binding energy ratio $\zeta_1$. Each group includes two important parameters: the envelope total to gravitational binding energy ratio $\zeta$, defined in Equation \ref{eqn:zeta}, and the envelope initial Mach number $\mathcal{M}$. We list the density and temperature of the $\dot{M}=5M_{\odot}/\rm{yr}$ models in columns 7 and 8. Columns 9 and 10 are the kinetic and radiation plus gas internal energy injection rates associated with the expanding envelope, respectively.}
    \label{tab:ejectapara}
\end{table*}

We investigate the dynamics and evolution of the envelope through a parameter survey assuming constant injection properties at the inner boundary. We explore a grid of three primary parameters: the ratio of radiation to gas internal energy, $\E/\eg\in\{0.2,0.4,0.8,1.6,3.2\}$; the ratio of the envelope's initial injection speed to the local escape speed, $\bar{v}_{\rm{ej}}=v_{\rm{ej}}/\sqrt{2G\M/\rin}\in\{0.3,0.4,0.5,0.6\}$; and the mass injection rate, $\dot{M}\in\{2.5,5,10\}M_{\odot}/\rm{yr}$, (see Table \ref{tab:ejectapara}). We observe that at the highest mass injection rate of $10M_{\odot}/\rm{yr}$, the total mass loss over 50 days reaches approximately $1.37M_{\odot}$, representing $9\%$ of the binary's total mass. As this mass loss progresses, the gravitational potential is expected to become shallower. Consequently, more sophisticated models that account for both the evolution of central gravity and the effects of self-gravity should be prioritized in future studies. Each model is designated by the identifier ``ExxVxxMxxx"; for example, model E16V03M025 corresponds to parameters $\E/\eg=1.6$, $\bar{v}_{\rm{ej}}=0.3$, and $2.5M_{\odot}/\rm{yr}$.

The combination of these parameters yields a range of initial Mach numbers, $\mathcal{M}=v_{\rm{ej}}/c_s$, which are listed in the corresponding segments of Table \ref{tab:ejectapara}. A significant subset of our models features initially subsonic envelopes, allowing us to characterize the acceleration phase from a quasi-hydrostatic state to an unbound outflow. We adopt a fixed duration for the expanding phase, $\Delta t=50$ days, and neglect the impact of self-gravity.

To gauge the potential of the envelope to escape the gravitational potential of the central object, we define three dimensionless parameters:
\begin{eqnarray}
    \zeta_1&=&\frac{\rin v_{\rm{ej}}^{2}/2}{G\M},    \label{eqn:zeta1}\\
    \zeta_2&=&\frac{\rin(v_{\rm{ej}}^{2}/2+\eg/\rho)}{G\M},   \label{eqn:zeta2}\\
    \zeta&=&\frac{\rin[v_{\rm{ej}}^{2}/2+(\eg+\E)/\rho]}{G\M}.   \label{eqn:zeta}
\end{eqnarray}

These variables compare different components of the specific energy to the specific gravitational binding energy. $\zeta_1$ serves as a conservative estimate, accounting only for the kinetic energy. $\zeta_2$ includes the gas internal energy (thermal and recombination); however, previous studies suggest that recombination energy alone may be inefficient at unbinding the envelope in the absence of radiation transport \citep{grichener2018,soker2018}. Consequently, $\zeta_2$ may not fully represent the escape capability in purely hydrodynamic scenarios. However, the inclusion of radiation transport and pressure requires a reevaluation of this efficiency. As noted in \citetalias{chen2024}, the radiative acceleration $a_{\rm{rad}}$ can exceed the gravitational acceleration $g$ in ionized regions. Therefore, we define $\zeta$ to incorporate the total energy density ($\E$ and $\eg$), serving as an optimistic measure of the envelope's unbound state. We anticipate that the synergy between radiation pressure and recombination energy plays a critical role in envelope ejection.

\section{Simulation results}\label{sec:results}

All the models are evolved for 500 days, as most dynamical phenomena have faded afterward.

\subsection{A typical envelope evolution}\label{sec:singleevolution}

\begin{figure*}
    \centering
    \includegraphics[width=\textwidth]{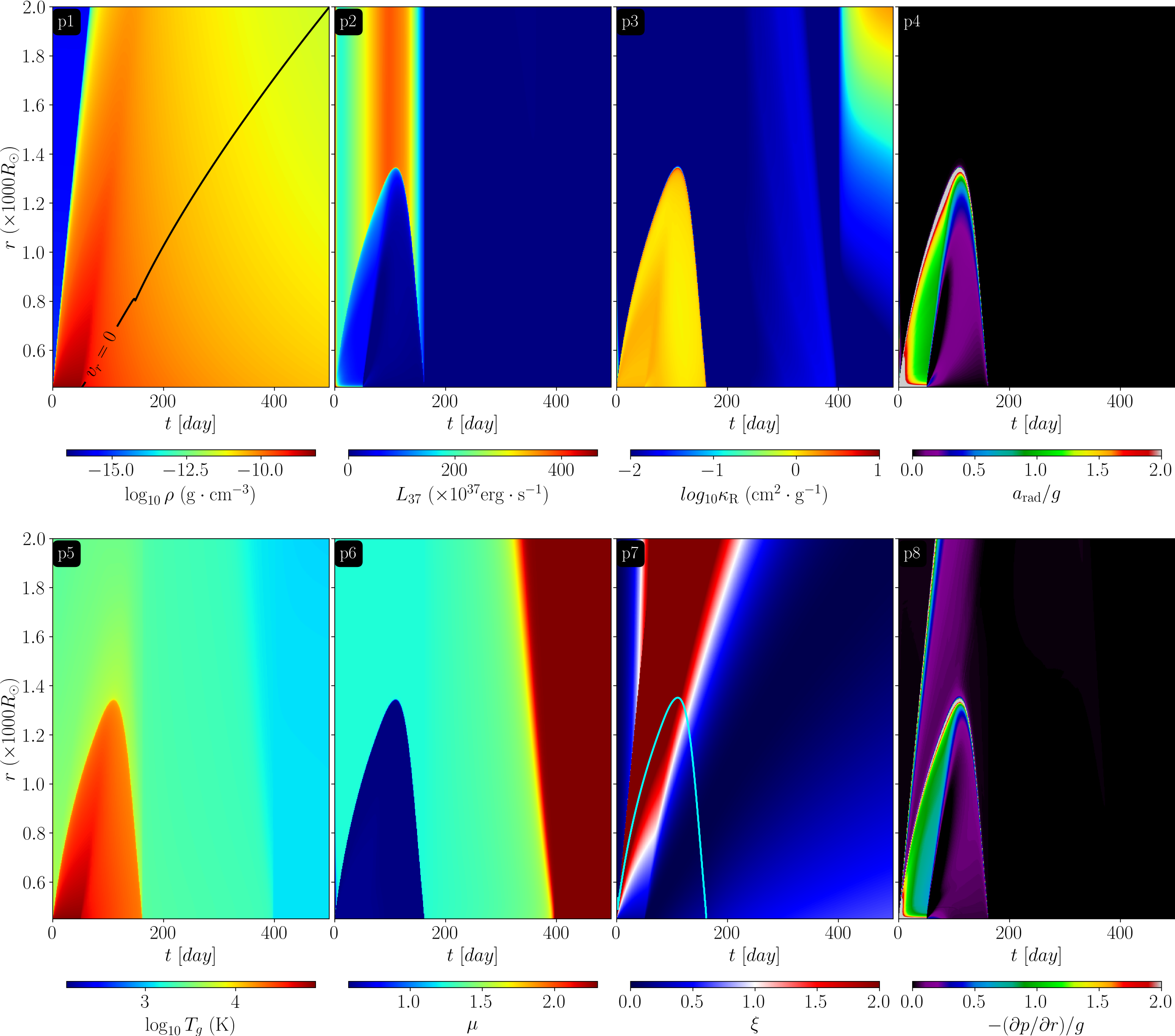}
    \caption{The evolution of model E16v03M050 ($\E/\eg=1.6$, $\bar{v}_{\rm{ej}}=0.3$, and $\dot{M}=5M_{\odot}/\rm{yr}$). In each panel, the x-axis is time in days, and the y-axis is the $r$ coordinate in $1000R_{\odot}$. From panels 1 to 8, they are the evolution of the density in \gcmc, local luminosity in $10^{37}\rm{erg}\cdot\rm{s}^{-1}$, Rosseland mean opacity in \cmsg, radiation pressure acceleration compared to the gravitational force, gas temperature in K, mean atomic weight, $\xi$ defined Equation \ref{eqn:xi} with the $\tg=6000$K isotherm, and negative pressure gradient compared to the gravitational force.}
    \label{fig:singleevolution}
\end{figure*}

The qualitative evolution of the envelope remains consistent across the surveyed parameter space. We select model E16V03M050 as a representative case, illustrating its evolution in Figure \ref{fig:singleevolution}.

Panel 1 displays the density evolution overlaid with the $v_{r}=0$ contour, which bifurcates the envelope into an outward-moving ejecta component and an inner fallback region. As detailed in Section \ref{sec:boundcond}, the inner boundary transitions to an inflow-only state following the injection phase; consequently, fallback material accretes onto the common envelope and effectively exits the computational domain, ceasing to influence the outer envelope.

Panel 2 depicts the local radiative luminosity, defined as $L_{37} = 4\pi r^{2}F_{r}/(10^{37}\,\rm{erg}\,\rm{s}^{-1})$, where $F_{r}$ represents the radiation flux. By cross-referencing Panel 2 with the temperature (Panel 5) and mean atomic weight $\mu$ (Panel 6), we observe that the local luminosity initially rises during expansion but drops precipitously upon hydrogen recombination. Correspondingly, the opacity (Panel 3) remains high in the ionized region due to bound-free and free-free transitions, but decreases by over two orders of magnitude in the neutral regime. Together, Panels 2 and 3 demonstrate that the transition to free-streaming radiation transport is spatially coincident with the recombination front. The rapid expansion and subsequent recession of this front---inferred from multi-band photometry of LRNe such as AT 2019zhd, AT 2020kog, and AT 2021blu \citep{pastorello2021a,pastorello2021b,pastorello2023}---are qualitatively reproduced here with a different timescale, consistent with the results in \citetalias{chen2024}. Notably, $L_{37}$ increases gradually beneath the recombination front due to radiative diffusion. Driven by the combination of high luminosity and high opacity, the ratio of radiative acceleration to gravity, $a_{\rm{rad}}/g$ (Panel 4), becomes substantial in this region. Wherever $a_{\rm{rad}}/g>1$, radiation pressure exerts a significant accelerating force on the envelope. Acceleration is also provided by the gas pressure gradient (Panel 8), which plays a relatively larger role in regimes where $\E/\eg \ll 1$. Finally, we define the parameter $\xi$ to quantify the kinetic capability of the gas to escape:
\begin{equation}
\label{eqn:xi}\xi=\frac{v_{r}^{2}r}{2G\M}.
\end{equation}
Panel 7 illustrates the evolution of $\xi$ alongside the $T_{\rm g}=6000$K isotherm. In our model, this isotherm closely tracks the recombination front, effectively demarcating the zone of strong radiation pressure acceleration. As indicated in Panel 7, the gas becomes increasingly unbound following the sustained acceleration from radiation and gas pressure gradients (see Section \ref{sec:radforce} for detailed discussion).

\subsection{Radiation and gas pressure gradients}\label{sec:radforce}

\begin{figure*}
    \centering
    \includegraphics[width=\textwidth]{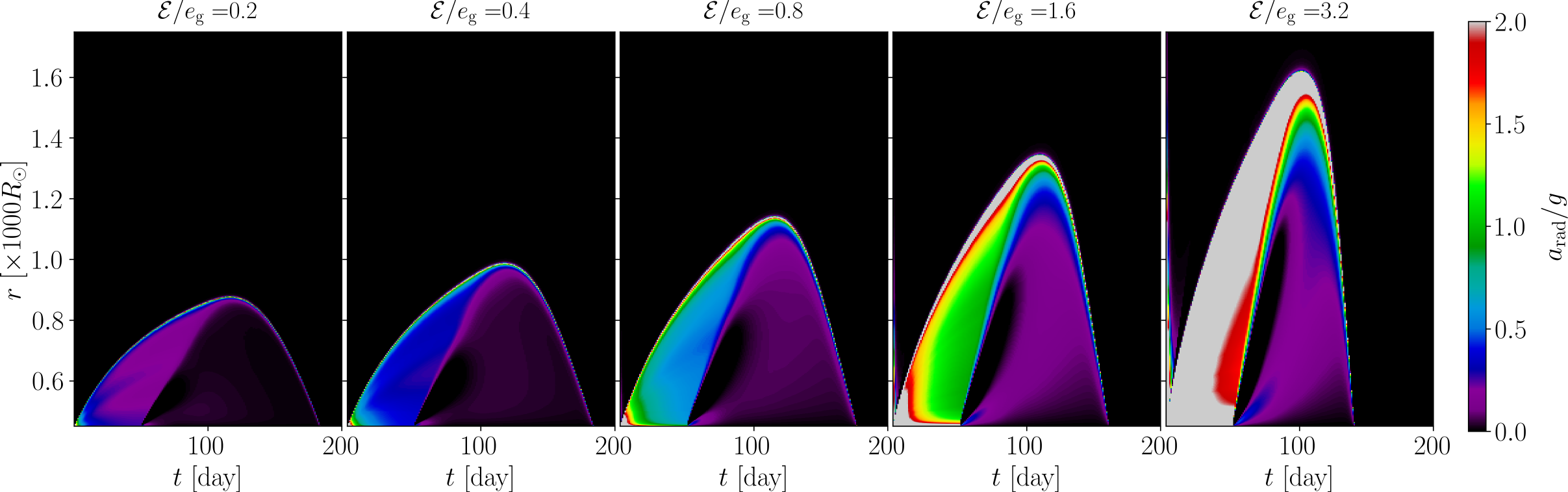}
    \caption{The evolution of $a_{\rm{rad}}$ of models with $\E/\eg\in\{0.2,0.4,0.8,1.6,3.2\}$, $\bar{v}_{\rm{ej}}=0.3$, and $\dot{M}=5M_{\odot}/\rm{yr}$. The grey and red color region indicates the radiation pressure dominated layer below the recombination front.}
    \label{fig:kipbetarad}
\end{figure*}
\begin{figure*}
    \centering
    \includegraphics[width=\textwidth]{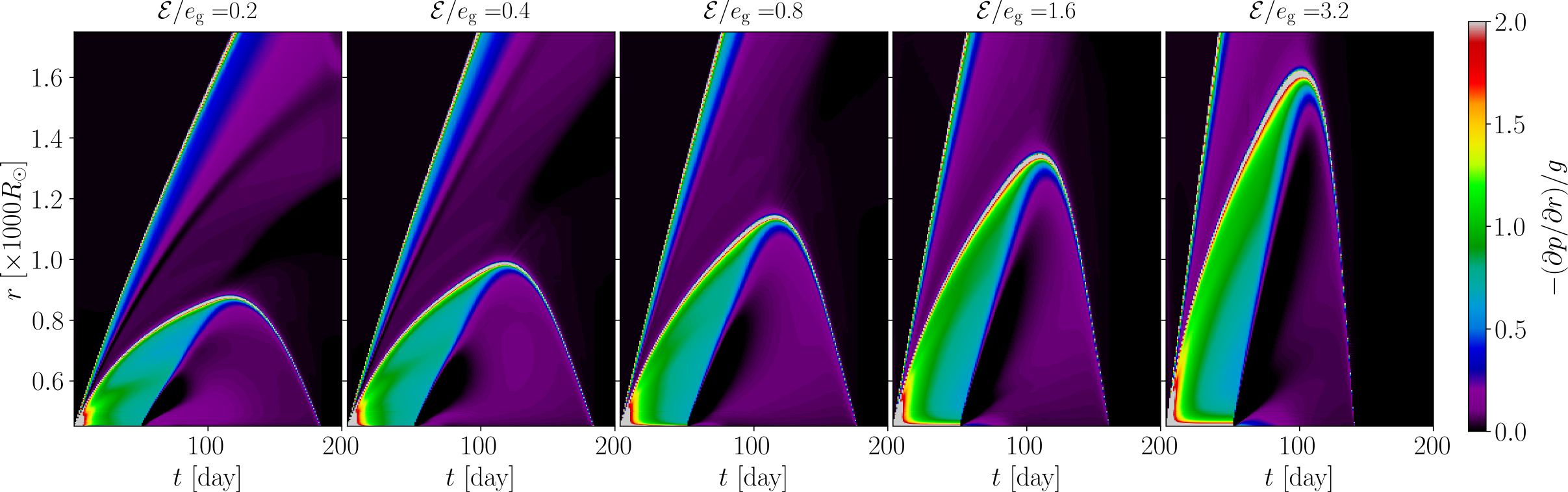}
    \caption{The evolution of $-\partial p/\partial r$ of models with $\E/\eg\in\{0.2,0.4,0.8,1.6,3.2\}$, $\bar{v}_{\rm{ej}}=0.3$, and $\dot{M}=5M_{\odot}/\rm{yr}$.}
    \label{fig:kipbetagas}
\end{figure*}
\begin{figure*}
    \centering
    \includegraphics[width=\textwidth]{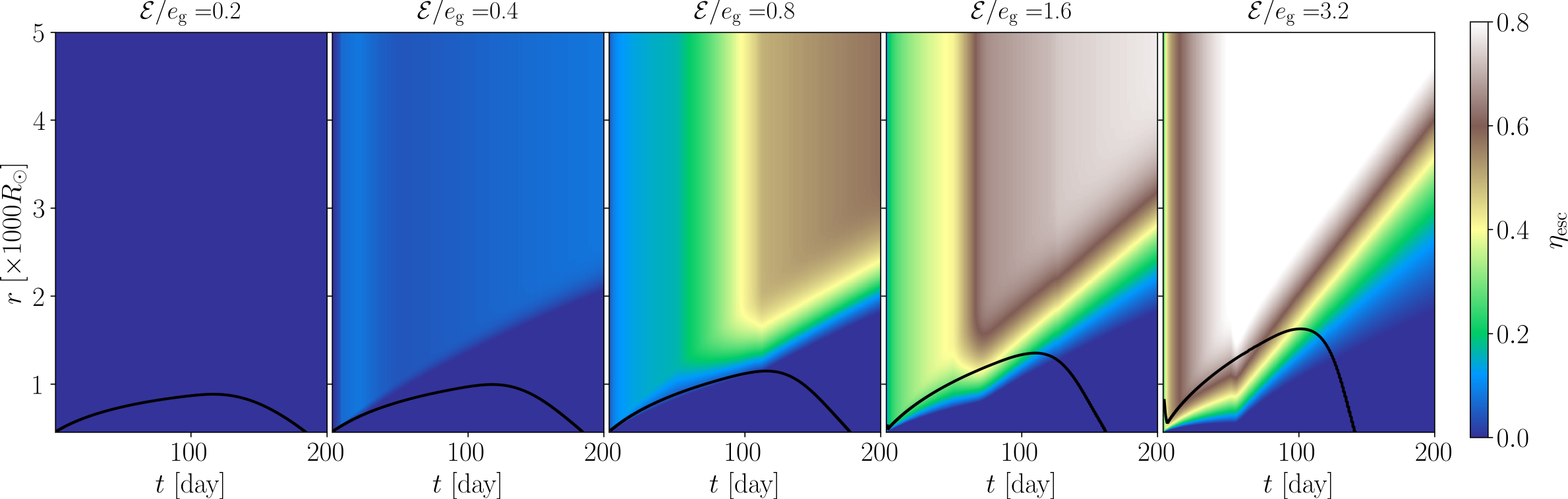}
    \caption{The evolution of $\eta_{\rm{esc}}$ of models with $\E/\eg\in\{0.2,0.4,0.8,1.6,3.2\}$, $\bar{v}_{\rm{ej}}=0.3$, and $\dot{M}=5M_{\odot}/\rm{yr}$. The black lines are the $\tg=6000$K isotherms, which almost coincide with the recombination fronts.}
    \label{fig:kipbetaeta}
\end{figure*}

We analyze the mechanics of envelope acceleration and unbinding as a function of the dimensionless parameter $\E/\eg$. Figures \ref{fig:kipbetarad} and \ref{fig:kipbetagas} show the evolution of $a_{\rm{rad}}/g$ and $-(\partial p/\partial r)/g$ of models with $\E/\eg\in\{0.2,0.4,0.8,1.6,3.2\}$, $\bar{v}_{\rm{ej}}=0.3$, and $\dot{M}=5M_{\odot}/\rm{yr}$. We observe a distinct contrast in the response of these forces: as $\E/\eg$ increases, $a_{\rm{rad}}$ rises significantly (by factors of several), whereas the gas pressure gradient remains relatively insensitive to the energy ratio. To quantify the ejection efficiency, we define the cumulative unbound mass fraction, $\eta_{\rm{esc}}(r)$, as:
\begin{equation}
    \eta_{\rm{esc}}(r)=\frac{\int_{\rin}^{r}dM_{\rm{esc}}(r)}{\int_{\rin}^{r}dM(r)},
\end{equation}
where
\begin{equation}
dM_{\rm{esc}}(r) = \begin{cases}
4\pi\rho r^{2}dr & \text{if } \xi > 1, \\
0 & \text{otherwise}.
\end{cases}
\end{equation}

Figure \ref{fig:kipbetaeta} displays the evolution of $\eta_{\rm{esc}}$ alongside the $T_{\rm g}=6000$K isotherms; the latter serve as a proxy for the recombination front. As $\E/\eg$ increases, a progressively larger fraction of the envelope is accelerated to an unbound state. Notably, due to the substantial enhancement of $a_{\rm{rad}}$ in high-energy models, the envelope material is able to reach the unbound state ($\xi > 1$) even within the deeper layers below the recombination front.

\subsection{The unbound mass fraction}

\begin{figure}
    \centering
    \includegraphics[width=\columnwidth]{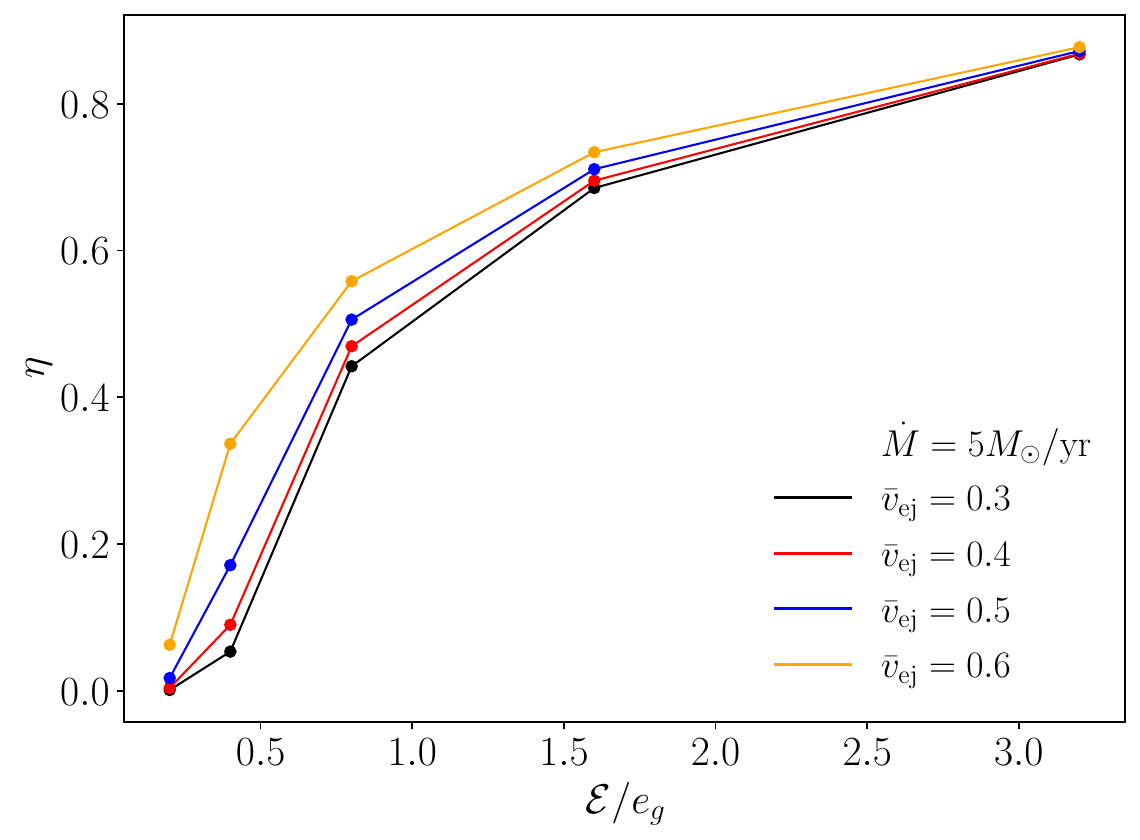}
    \includegraphics[width=\columnwidth]{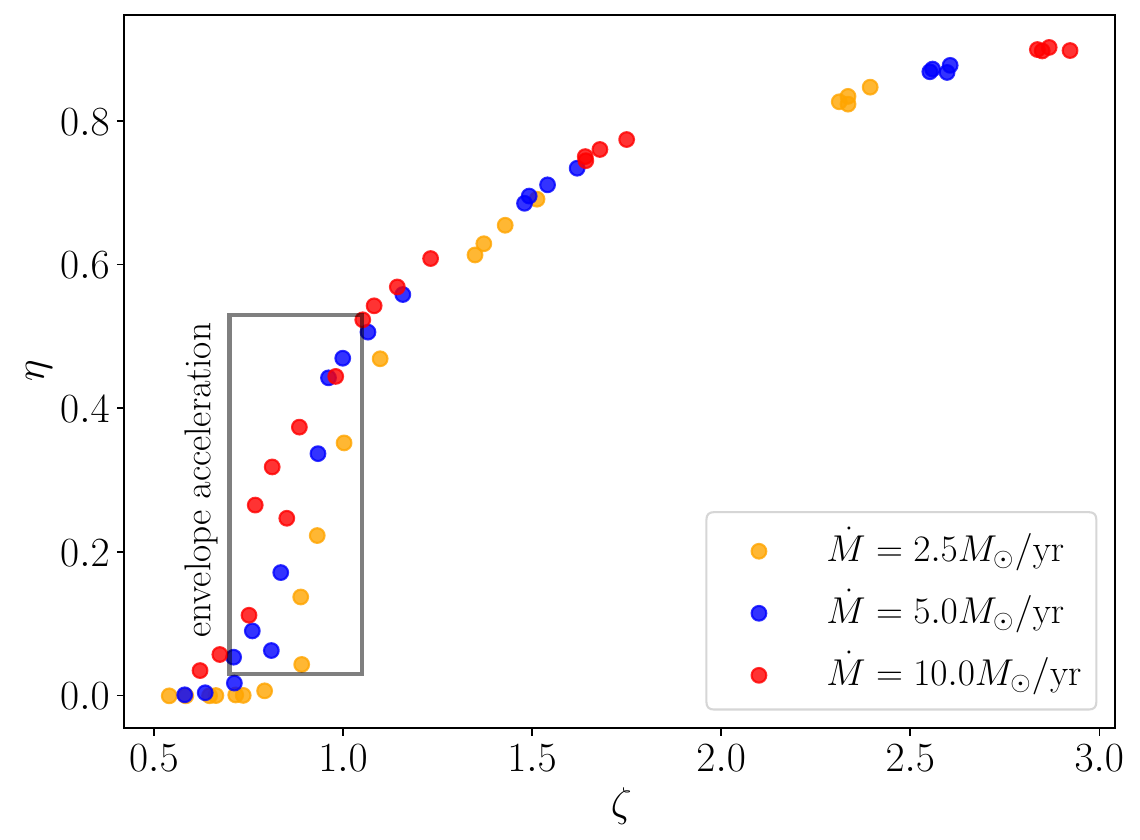}
    \caption{Upper panel: the unbound mass fraction $\eta$ v.s. different $\E/\eg$ and $\bar{v}_{\rm{ej}}$ for the $\dot{M}=5M_{\odot}/\rm{yr}$. Bottom panel: $\eta$ v.s. $\zeta$ of models with different $\dot{M}$. There is potentially a nonlinear relation between $\eta$ and $\zeta$.}
    \label{fig:radforce}
\end{figure}

We define the total unbound mass fraction, $\eta$, as the value of the cumulative distribution function evaluated at the outer boundary at the end of the simulation: $\eta \equiv \eta_{\rm{esc}}(\rout \mid t=t_{\rm{final}})$. As illustrated in the upper panel of Figure \ref{fig:radforce}, $\eta$ correlates positively with both $\E/\eg$ and $\bar{v}_{\rm{ej}}$, a trend consistent with physical expectations.

We further analyze the ejection efficiency using the dimensionless parameter $\zeta$, which represents the ratio of total envelope energy to gravitational binding energy. The bottom panel of Figure \ref{fig:radforce} indicates that envelope unbinding is inefficient when $\zeta<0.7$. However, in the regime where $\zeta\in[0.7, 1.05]$ (highlighted region), acceleration becomes effective; notably, we find that a higher $\dot{M}$ results in a higher $\eta$. This enhancement arises because the high optical depth of a dense envelope traps radiation, allowing the gas to acquire significant momentum. Conversely, in tenuous envelopes, radiation decouples and escapes more easily, contributing less to the pressure gradient—a scenario likely characteristic of the simulations by \citet{lau2025}.

The relationship between $\eta$ and $\zeta$ is nonlinear and sensitive to $\bar{v}_{\rm{ej}}$, $\dot{M}$, and $\M/\rin$. A broader parameter survey is required to fully constrain the relationship between orbital energy deposition and envelope unbinding, which could provide critical quantitative insights into the canonical $\alpha_{\rm{CE}}$ formalism \citep{webbink1984,demarco2011,hirai2022}.

\subsection{Light curves}\label{sec:lightcurve}

\begin{figure}
    \centering
    \includegraphics[width=\columnwidth]{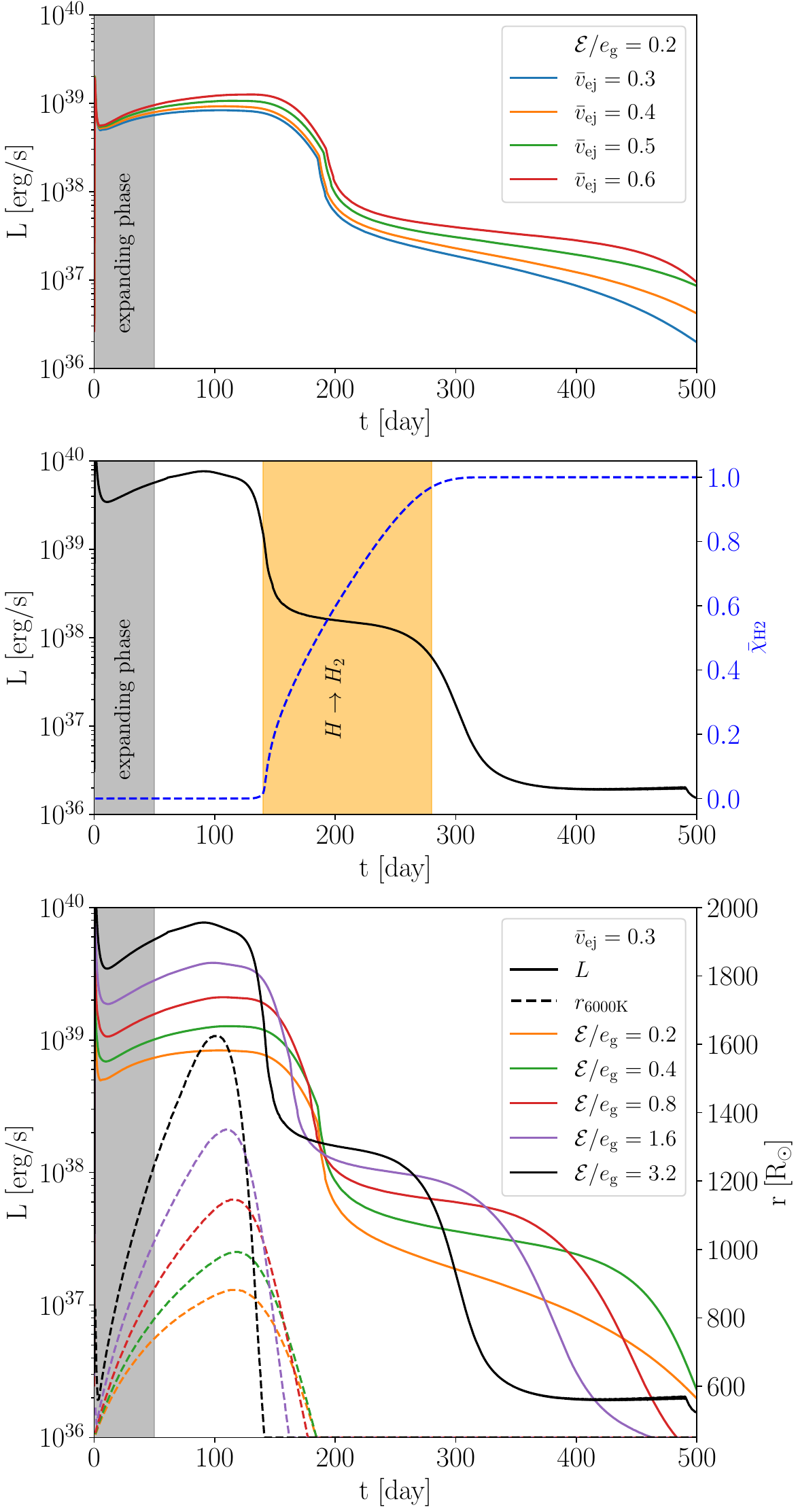}
    \caption{The grey color shaded regions highlight the envelope expanding phase. Top panel: light curves of models  with $\E/\eg=0.2$, $\bar{v}_{\rm{ej}}\in\{0.3,0.4,0.5,0.6\}$, and $\dot{M}=5M_{\odot}/\rm{yr}$. Middle panel: light curves of model E32V03M050. The secondary y-axis shows the total mass fraction of $\ce{H2}$. Bottom panel: light curves of models with $\bar{v}_{\rm{ej}}=0.3$, $\E/\eg\in\{0.2,0.4,0.8,1.6,3.2\}$, and $\dot{M}=5M_{\odot}/\rm{yr}$. The secondary y-axis shows the radii of $\tg=6000$K isotherms, which track the recombination front.}
    \label{fig:lightcurves}
\end{figure}

We examine how the parameters $\E/\eg$ and $\bar{v}_{\rm{ej}}$ influence the bolometric light curve, defined by the luminosity at the outer boundary:
\begin{equation}
\label{eqn:luminosity}L=4\pi\rout^{2}F_{r}(\rout).
\end{equation}

The top panel of Figure \ref{fig:lightcurves} displays the light curves for models with fixed $\E/\eg=0.2$ and $\dot{M}=5\,M_{\odot}\,\rm{yr}^{-1}$, varying $\bar{v}_{\rm{ej}}$ across $\{0.3,0.4,0.5,0.6\}$. As expected, envelopes with higher $\bar{v}_{\rm{ej}}$ expand more rapidly. Despite these temporal shifts, the overall morphology of the four light curves remains similar, suggesting that $\bar{v}_{\rm{ej}}$ is not the primary determinant of the light curve shape.

The middle panel illustrates the evolution of model E32V03M050. A distinct secondary plateau appears between 140 and 280 days. This feature is powered by the recombination of \ce{H} into \ce{H2}, which releases latent heat into the ejecta. To demonstrate this connection, we plot the mass fraction of molecular hydrogen relative to the total hydrogen mass:
\begin{equation}
\bar{\chi}_{\ce{H2}}=M_{\ce{H2}}/(XM),
\end{equation}
where $M_{\ce{H2}}$ is the total mass of \ce{H2}, $M$ is the total mass in the computational domain, and $X$ is the hydrogen mass fraction. The onset of the secondary plateau coincides with the rise of $\bar{\chi}_{\ce{H2}}$. Physically, the recombination of \ce{H} is often associated with dust formation. If the ejecta is sufficiently massive, the release of this binding energy may power emission in the infrared band, potentially making the diffused ejecta resemble the ``SPRITEs" (unusual infrared transients) observed by the Spitzer telescope \citep{kasliwal2017}.

The bottom panel compares light curves (solid lines) and $T_{\rm g}=6000$K isotherms (dashed lines) for models with $\bar{v}_{\rm{ej}}=0.3$ and $\dot{M}=5\,M_{\odot}\,\rm{yr}^{-1}$, across varying energy ratios $\E/\eg\in\{0.2,0.4,0.8,1.6,3.2\}$. Unlike the variations in velocity, varying $\E/\eg$ exhibits significant morphological diversity in light curve profiles. We observe a trend where high $\E/\eg$ envelopes achieve higher peak luminosities but fade more rapidly. This behavior arises because: (1) stronger radiation pressure accelerates the envelope more efficiently; (2) the higher $\E$ maintains ionization over a larger volume, pushing the recombination front outward; and (3) the rapid expansion leads to a faster density drop, allowing radiation to escape the system more easily. The resulting diversity in these light curves is thus driven by the interplay between ejecta kinematics and the thermodynamics of the recombination front.

In more realistic scenarios, the common envelope ejecta can be aspherical or even highly asymmetric, as demonstrated by the 3D hydrodynamical simulations mentioned previously. Such asymmetric ejecta may be associated with varying ratios of $\E/\eg$. Depending on the viewing angle, these asymmetries likely contribute to the observed diversity in light curves. Consequently, multi-dimensional radiation hydrodynamic modeling of LRNe is essential to improve the predictive power of the transition from CEE to LRNe \citep{kirilov2025}.

\section{Conclusions and Discussions}\label{sec:conclusion}

We investigated the envelope unbinding mechanisms during the plunge-in phase of CEE using a 1D RHD model evolved with the code {\tt Guangqi}. Through a parameter survey, we assessed the influence of $\E/\eg$, $\bar{v}_{\rm{ej}}$, and $\dot{M}$ on the ejection process, covering both subsonic and supersonic expansion regimes. Error analysis (Appendix \ref{app:energyconservation}) indicates that total energy may be underestimated in radiation-dominated or subsonic states; however, our simulations demonstrate robust convergence during the dynamical phase (Appendix \ref{app:convergence}). We draw three primary conclusions from this work:
\begin{enumerate}
\item Radiation Pressure Acceleration: High values of $\E/\eg$ drive significant acceleration (Figure \ref{fig:kipbetarad}), particularly in the layer immediately beneath the recombination front. In this region, the product of the local luminosity (enhanced by diffusion) and opacity (dominated by bound-free and free-free transitions) is maximal. This mechanism leads to a larger unbound mass fraction, $\eta$ (Section \ref{sec:radforce}), and a higher peak luminosity in the resulting light curves (Section \ref{sec:lightcurve}).
\item $\eta-\zeta$ Relation: We identify a potentially nonlinear relationship between the energy parameter $\zeta$ and the unbound fraction $\eta$, which may be modulated by $\bar{v}_{\rm{ej}}$, $\dot{M}$, and $\M/\rin$ (Figure \ref{fig:radforce}). This relationship offers quantitative insights that could help refine the standard $\alpha_{\rm{CE}}$ formalism.
\item Recombination Plateaus: The recombination of atomic hydrogen to molecular hydrogen (\ce{H} $\to$ \ce{H2}) releases latent heat during late-time evolution, powering a plateau phase in the bolometric light curve (Figure \ref{fig:lightcurves}).
\end{enumerate}

Physically, the unbinding process likely involves a large amount of energy deposition deep inside the envelope, accompanied by convective dissipation. A fully self-consistent model requires an inner boundary condition directly coupled to the binary inspiral to accurately drive gas and radiation pressure, convective transport, and accommodate the fallback envelope. Developing such a boundary condition represents a critical avenue for future work. Finally, we note the inherent limitations of the 1D approach in capturing angular momentum transport and geometric asymmetries, both of which are fundamental to CEE. Future 2D and 3D parameter surveys will be essential to quantify the impact of angular momentum transfer and asymmetric outflows on the unbinding efficiency and observational signatures.

\section*{}

The author thanks the anonymous referee who has provided incisive suggestions that improved the quality of this paper. This work is financially supported by the National Science Foundation of China under grants No. 12103028, 12342501, 12473030, Tsinghua University Dushi Program at Tsinghua University. The Center of High Performance Computing at Tsinghua University provided the computational resources. Z.C. thanks the International Centre of Supernovae (ICESUN), Yunnan Key Laboratory of Supernova Research for providing a vibrant environment for discussion.

\software{{\tt Guangqi} \citep{chen2025}, {\tt Matplotlib} \citep{hunter2007}, {\tt Petsc} \citep{petsc-efficient,petsc-user-ref}}

\appendix
\section{Energy conservation}\label{app:energyconservation}

Because the gravitational force and a part of the radiation pressure are treated as explicit source terms in the energy equation, and the implicit radiation transport solver cannot conserve energy to machine precision, it is necessary to examine the energy conservation of our simulations. Dimensional analysis shows that \citep{colombo2019},
\begin{equation}\label{eqn:approximation}
    \nabla\cdot\mathcal{P}_{r}\approx-\frac{\rho\kr\F}{c},
\end{equation}
when $v_{\text{gas}}/c\ll1$. Because this condition is satisfied in all of our simulations, we analyze the energy conservation of our simulations by substituting Equation \ref{eqn:approximation} into \ref{eqn:hydroenergy}, and obtain
\begin{eqnarray}\label{eqn:hydroenergy_approx}
    \pdv{E}{t}+\nabla\cdot[(E+p)\vel]&=&- \vel\cdot(\nabla\cdot\mathcal{P}_{r}+\rho\nabla\phi)+\G.\quad\quad
\end{eqnarray}
Adding Equation \ref{eqn:hydroenergy_approx} and \ref{eqn:radenergy}, we obtain,
\begin{equation}
    \pdv{(E+\E+\rho\phi)}{t}+\nabla\cdot[(E+p+\E+\mathcal{P}_{r}+\rho\phi)\cdot\vel+\F]=0,
\end{equation}
where we have used $\nabla\cdot(\rho\vel\phi)=\rho\vel\cdot\nabla\phi+\phi\nabla\cdot(\rho\vel)$ and the mass conservation equation.

We examine the energy conservation by calculating the energy balance at the end of the simulations. The error of the energy $E_{\rm{err}}$ is calculated by,
\begin{eqnarray}
    E_{\rm{err}}&=&ET_{\rm{final}}-ET_{\rm{init}}+\int[F(\rin)-F(\rout)]dt\quad  \label{eqn:energyerror}\\
    ET&=&\int_{V}(E+\E+\rho\phi)dV    \\
    F(r)&=&\int_{S}[(E+p+\rho\phi+\E+\mathcal{P}_{r})\cdot\vel+\F]\cdot\mathbf{dS}
\end{eqnarray}
where $ET$ is the total energy in the computational domain, and $F(r)$ is the area integrated energy flux at radius $r$. Because $ET_{\rm{init}}$ is usually 5 orders of magnitude smaller than $ET_{\rm{final}}$, it is negligible. At the inner boundary, $\F(\rin)=0$ is determined by the boundary condition, and the environment at $\rin$ is always optically thick, thus, $\mathcal{P}_{r}\cdot\vel\approx\E\vel/3$. Combining the radiation pressure term with the radiation advection term, we get $4\E\vel/3=4\E v_{\rin}\hat{r}/3$, where $v_{\rin}$ denotes the velocity at $r=\rin$. At the outer boundary, the environment is always set to optically thin, therefore $|\mathcal{P}_{r}\cdot\vel|\approx|\E v_{\rout}|\ll\E c\approx F_{r}$ is negligible compared to the radiation transport. We further define two energies and a relative error by,
\begin{eqnarray}
    \Delta_{N}E_{\rm{hyd}}&=&4\pi\rin^{2}\int_{0}^{N\text{day}}[(E+p)v_{\rin}]dt\label{eqn:Ehyd}\\
    \Delta_{N}E_{\rm{rad}}&=&4\pi\rin^{2}\int_{0}^{N\text{day}}\frac{4\E v_{\rin}}{3}dt\label{eqn:Erad}\\
    \delta E_{\rm{err}}&=&\frac{E_{\rm{err}}}{\Delta_{50}E_{\rm{hydro}}+\Delta_{50}E_{\rm{rad}}}.\label{eqn:deltae2} 
\end{eqnarray}

\begin{figure}
    \centering
    \includegraphics[width=\columnwidth]{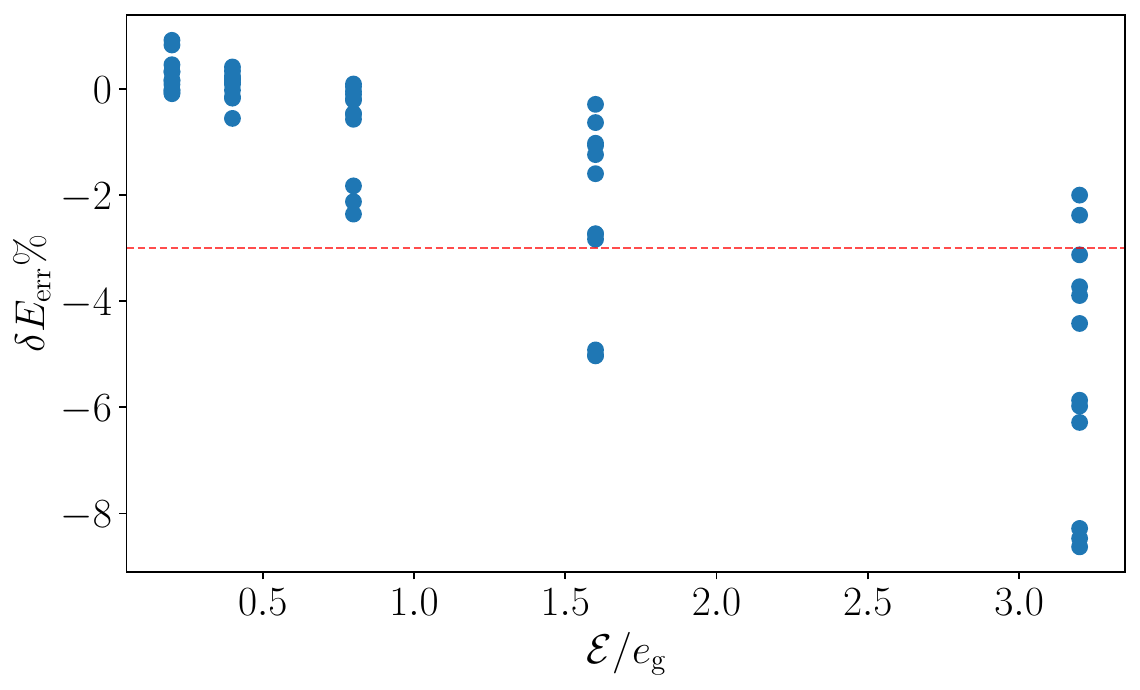}
    \includegraphics[width=\columnwidth]{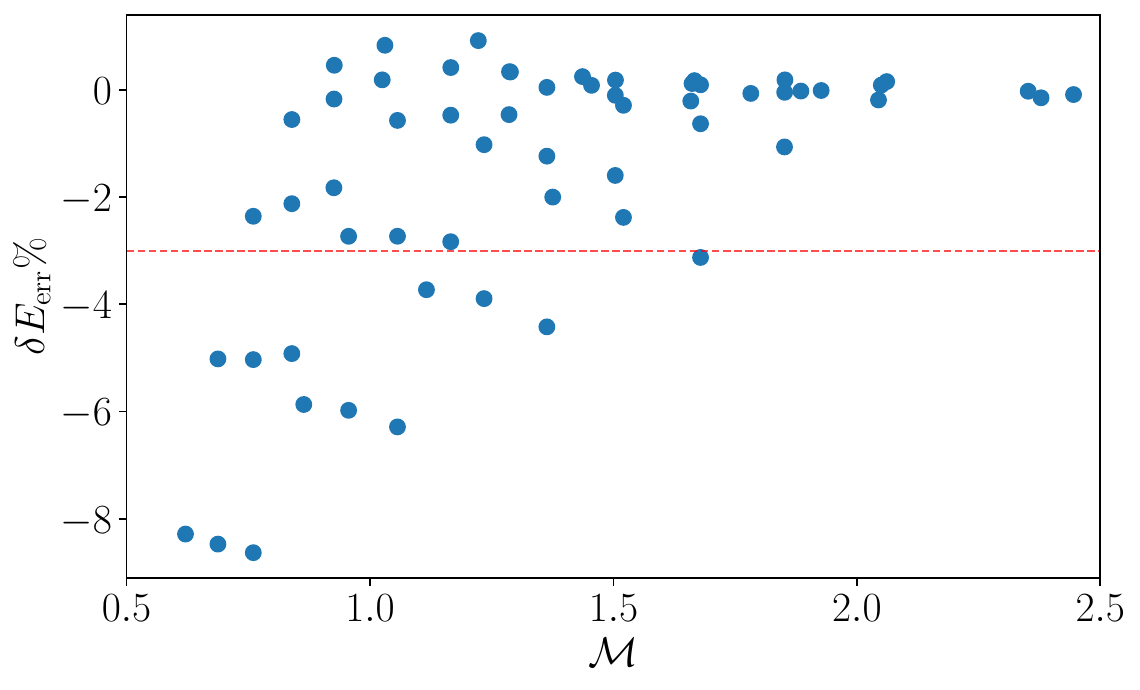}
    \caption{Relative energy errors of all the 60 simulations in Table \ref{tab:ejectapara}. Top panel: $\delta E_{\rm{err}}$\% v.s. $\E/\eg$. Bottom panel: $\delta E_{\rm{err}}$\% v.s. $\mathcal{M}$. The red lines indicate $\delta E_{\rm{err}}=-3\%$.}
    \label{fig:error}
\end{figure}

Figure \ref{fig:error} shows $\delta E_{\rm{err}}$ v.s. $\E/\eg$ and $\mathcal{M}$. Our model is quite accurate ($|\delta E_{\rm{err}}|\le3$\%) for the envelopes with $\E/\eg<1$ or $\mathcal{M}>1.5$. For the radiation-dominated or subsonic envelope, the error can grow to $\sim-8$\%. The minus sign indicates that our model is more likely to underestimate the unbound mass or luminosity. Further improving the accuracy is a future research direction.

Finally, we note that the total injected energy $\Delta_{50}E_{\rm{hyd}}+\Delta_{50}E_{\rm{rad}}$ is in the range of $[10^{47},3\times10^{47}]$ erg for the $\dot{M}=5M_{\odot}/\rm{yr}$ models, which is much smaller than the gravitational energy released ($\sim10^{48}$ erg) by a typical massive binary during the plunge-in phases \citep{fragos2019}. Thus, a full CEE of a similar kind of binary could involve more envelope unbinding and a brighter light curve.

\section{Numerical convergence}\label{app:convergence}

\begin{figure}
    \centering
    \includegraphics[width=\columnwidth]{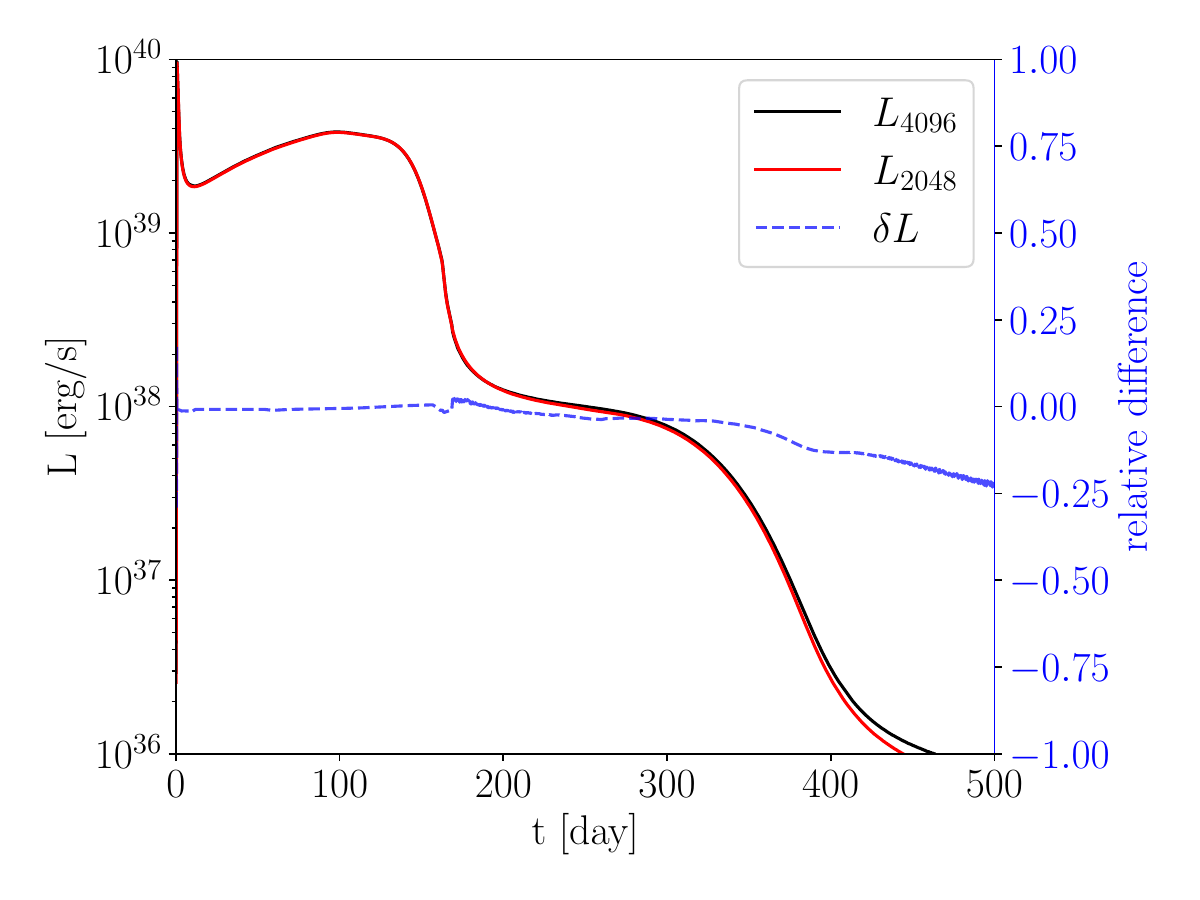}
    \caption{The luminosity $L$ (Equation \ref{eqn:luminosity}) of model E16V03M050. The black and red lines are the results from the $N_{x}=4096$ and $N_{x}=2048$ base resolution simulations, respectively. The blue dashed line is their relative difference, $\delta L=(L_{4096}-L_{2048})/L_{4096}$.}
    \label{fig:convergence}
\end{figure}

The RHD equations, coupled with a complex EoS, constitute a highly nonlinear and multiscale system. To ensure numerical stability and accuracy, we employ sub-cycling for the radiation transport solver, characterized by time-step control parameters $n_{\rm{sub}}=6$ and $q_{t}=1.4$.

To validate numerical convergence, we utilize the outer boundary luminosity, $L$ (Equation \ref{eqn:luminosity}), as a diagnostic metric, as it is sensitive to both grid resolution and parameter changes. We select model E16V03M050 as a representative test case. Characterized by a high $\E/\eg$ ratio and a subsonic envelope (Table \ref{tab:ejectapara}), this model is susceptible to larger numerical errors (see Figure \ref{fig:error}), making it a rigorous candidate for convergence testing.

We perform a resolution study by comparing the baseline resolution ($N_{x}=2048$) against a high-resolution run ($N_{x}=4096$). Doubling the spatial resolution results in a commensurate reduction of the hydrodynamic and radiation timesteps. Figure \ref{fig:convergence} presents the light curves from both simulations alongside their relative errors. The relative error remains negligible for the first 400 days---a duration that fully encompasses the critical phases of envelope acceleration and hydrogen recombination.

Beyond 400 days, the relative difference increases as the ejecta reaches large radii. This divergence may be attributed to the stretched grid, which results in coarser spatial resolution in the outer computational domain. However, since the primary dynamical events occur earlier, this late-time resolution drop does not affect our core conclusions. Our model successfully resolves the most intense dynamical phases of the expanding envelope, and convergence tests for other models yield qualitatively similar results.

\bibliography{cee1d}{}
\bibliographystyle{aasjournal}

\end{CJK*}
\end{document}